\documentclass[fleqn,usenatbib]{mnras}

\usepackage{newtxtext,newtxmath}

\usepackage[T1]{fontenc}
\usepackage{ae,aecompl}

\usepackage{graphicx}	
\usepackage{amsmath}	
\usepackage{amssymb}	
\usepackage{gensymb}
\usepackage{ragged2e}
\title[Early evolution of dwarfs]{On the early evolution of Local Group dwarf galaxy types: star formation and supernova feedback}
	
\author[J. R. Bermejo-Climent et al.]{Jos\'e R. Bermejo-Climent$^{1,2,3,4}$\thanks{E-mail: bermejo@iasfbo.inaf.it},
Giuseppina Battaglia$^{3,4}$, Carme Gallart$^{3,4}$, \newauthor Arianna Di Cintio$^{3,4,5}$, Chris B. Brook$^{3,4}$, Luis Cicu\'endez$^{3,4}$, Matteo Monelli$^{3,4}$, \newauthor Ryan Leaman$^{6}$, Lucio Mayer$^{7}$, Jorge Pe\~narrubia$^{8}$ and Justin I. Read$^{9}$
\\
$^{1}$INAF/OAS Bologna, Via Piero Gobetti 101, Area della Ricerca CNR/INAF, I-40129 Bologna, Italy \\
$^{2}$INFN, Sezione di Bologna, Via Irnerio 46, I-40126 Bologna, Italy\\
$^{3}$Instituto de Astrof\'isica de Canarias, C/Via Lactea, s/n, 38205 La Laguna, Tenerife, Spain\\
$^{4}$Departamento de Astrof\'isica, Universidad de La Laguna, 38206 La Laguna, Tenerife, Spain\\
$^{5}$Leibniz Institute for Astrophysics Potsdam (AIP), An der Sternwarte 16, D-14482 Potsdam, Germany\\
$^{6}$Max-Planck Institut f\"ur Astronomie, K\"onigstuhl 17, D-69117 Heidelberg, Germany\\
$^{7}$Center for Theoretical Astrophysics and Cosmology, Institute for Computational Science, University of Zurich,\\ Winterthurerstrasse 190, CH-8057 Zurich, Switzerland\\
$^{8}$Institute for Astronomy, University of Edinburgh, Blackford Hill, Edinburgh EH9 3HJ, UK\\
$^{9}$Department of Physics, University of Surrey, Guildford, GU2 7XH, UK\\
}

\date{Accepted 2018 June 20. Received 2018 June 15; in original form 2018 March 28}

\pubyear{2018}

\begin{document}
\label{firstpage}
\pagerange{\pageref{firstpage}--\pageref{lastpage}}
\maketitle

\begin{abstract}

According to star formation histories (SFHs), Local Group dwarf galaxies can be broadly classified in two types: those forming most of their stars before $z=2$ ({\it fast}) and those with more extended SFHs ({\it slow}). The most precise SFHs are usually derived from deep but not very spatially extended photometric data; this might alter the ratio of old to young stars when age gradients are present. Here we correct for this effect and derive the mass formed in stars by $z=2$ for a sample of 16 Local Group dwarf galaxies. We explore early differences between {\it fast} and {\it slow} dwarfs, and evaluate the impact of internal feedback by supernovae (SN) on the baryonic and dark matter (DM) component of the dwarfs. {\it Fast} dwarfs assembled more stellar mass at early times and have larger amounts of DM within the half-light radius than {\it slow} dwarfs. By imposing that {\it slow} dwarfs cannot have lost their gas by $z=2$, we constrain the maximum coupling efficiency of SN feedback to the gas and to the DM to be $\sim$10\%. We find that internal feedback alone appears insufficient to quench the SFH of {\it fast} dwarfs by gas deprivation, in particular for the fainter systems. Nonetheless, SN feedback can core the DM halo density profiles relatively easily, producing cores of the sizes of the half-light radius in {\it fast} dwarfs by $z=2$ with very low efficiencies. Amongst the ``classical" Milky Way satellites, we predict that the smallest cores should be found in Draco and Ursa Minor, while Sculptor and Fornax should host the largest ones.
\end{abstract}




\section{Introduction}

Dwarf galaxies are the smallest and most numerous galaxies in the Universe. In the range of absolute magnitudes $M_V >$ -15, they are typically classified into dwarf spheroidals (dSphs), gas-poor and passively evolving; dwarf irregulars (dIrrs), which are gas-rich and star-forming systems (e.g. \citealt{mateo98}); and transition types (dTs) with intermediate properties between dSphs and dIrrs. The existence of different kinds of dwarf galaxies has opened a question that is still unsolved \citep{skillman}: are the current properties of dwarf galaxies  a result of their evolution or were they imprinted during their early assembly?

The star formation histories (SFHs) are an exquisite tool that allows us to reconstruct the lifetime evolution of galactic systems. \citet{gallart} compared {\it accurate} literature SFHs for a sample of 18 Local Group (LG) dwarf galaxies, selecting only those derived from deep photometric data reaching down to below the oldest main sequence turnoff (oMSTO), and established an alternative classification of dwarf galaxies based on their lifetime evolution: they were divided in {\it fast} dwarfs, those that formed the majority of their stellar component early on (before $z \simeq 2$), and  {\it slow} dwarfs, that only formed a small fraction of their stars at early times and continued forming stars for almost a Hubble time. The proposed dichotomy is not equivalent to the commonly adopted, traditional one: all the {\it fast} dwarfs are dSphs or transition types; however, not all the {\it slow} types are dIrrs: e.g. Milky Way (MW) satellites such as the Carina, Fornax and Leo~I dSphs can be classified as having {\it slow} SFHs.

In \citet{gallart} the SFHs were normalized and compared in a relative way. However, many of these SFHs are derived from photometric data that are deep but do not cover the entire stellar component. This, and the presence of age gradients in the dwarfs' stellar population (e.g. \citealt{bat6}, \citealt{bat12}, \citealt{bat12b}, \citealt{hi13}, \citealt{delpino15}) can result in an underestimation of the star formation at early times, since usually older stellar populations present more extended spatial distributions. Here we extend the analysis by \citet{gallart} correcting for the missing spatial coverage of the ancient stars for a similar sample of LG dwarf galaxies. This allows us to integrate the absolute amount of mass formed into stars up to $z \simeq 2$ ($\sim $10 Gyr ago) in order to learn more about possible early differences between {\it fast} and {\it slow} types. We focus on the stellar mass formed up to redshift $z=2$ ($>$10 Gyr ago), $M_{\star, z>2}$, because at lower redshift the two types are already very distinct, in that most {\it fast} dwarfs have experienced no or very little star formation.

Much of the theoretical research about the evolution of dwarf galaxies focuses on answering how could dSphs have lost their gas. There have been proposed many environmental mechanisms for the gas removal, such as ram-pressure or tidal stripping by a massive central halo (\citealt{grebel03}, \citealt{mayer06}, \citealt{mayer10},  \citealt{gatto}) and the effects of an ionizing cosmic UV background (\citealt{efsta}, \citealt{bullock}, \citealt{salvadori}, \citealt{sawala10}). The importance of environmental effects is supported by the existence of the observed morphology-density relations in galaxy groups: dwarfs with different gas content are preferentially found in different environments, with dSphs usually inhabiting denser locations as the neighborhood of a large galaxy like the MW or M31.

Another explored scenario for the gas removal is the internal feedback and the gas ejection through supernova-driven outflows (\citealt{dekel}, \citealt{maclow}, \citealt{salvadori}, \citealt{sawala10}, among others). Our derivation of the stellar mass formed up to $z \simeq 2$ allows us to estimate the amount of energy injected by the supernovae (SN) to the inter-stellar medium (ISM) up to this redshift, and to quantify if the stellar feedback is enough or not to remove the gas and quench the star formation on {\it fast} dwarfs at early times. This can be done by estimating the competing effect of the gravitational potential of the dark matter (DM) halo versus the SN energy injected. Since kinematic measurements can essentially determine the dynamical mass within the spatial extent of the kinematic tracer, we use abundance-matching (AM) relations to link the total DM halo mass to the stellar mass of each galaxy (e.g. \citealt{behroozi}, \citealt{moster}, \citealt{brook14}). 

The calculation of the amount of SN feedback in dwarf galaxies can also provide information about whether this energy could or not change the DM halo density profile of the different kinds of dwarfs. It has been proposed (\citealt{nav96}, \citealt{read05}, \citealt{pontzen}) that the feedback energy coupled to the gas can subsequently modify the DM distribution by gravitational effects. This would solve the so-called `cusp/core' problem, that is the mismatch between the observed mass profiles, consistent with homogeneous-density `cores' (e.g. \citealt{flores}, \citealt{moore}, \citealt{kuzio}, \citealt{bat08}, \citealt{blok}, \citealt{walker11}, \citealt{amorisco}) and the cosmological N-body simulations suggesting that if gravitational interactions between cold dark matter (CDM) particles dominate the structure formation, the DM density profiles are characterized by centrally divergent `cusps' (\citealt{dubinski}, \citealt{nfw96}). 

A similar long-standing tension between observations of the nearby Universe and the standard cosmological model is the ``missing satellites" problem: $\Lambda$CDM simulations produce more DM haloes than observed galaxies, also in Local Group-like environments (\citealt{klypin98}, \citealt{moore98}). The mismatch can be explained with baryonic physics and likely involves supernova feedback, an ionising UV background, tidal stripping and possibly cusp-core transformations all working together in concert (e.g. \citealt{maccio10}, \citealt{zolotov12}, \citealt{wetzel}, \citealt{sawala16b}); however, assuming a `cuspy' NFW profile \citep{nfw}, kinematic measurements suggest that the smallest galaxies would live in the smallest haloes, leaving some inhabited haloes that would not be small enough to prevent star formation, this is called the `too big to fail' problem \citep{bk11,gk14}. The solution within the CDM paradigm to match these galaxies with larger DM haloes is the presence of cores that would explain the lower measured velocity dispersions \citep{brooks14,brook15,brook15b}.
The connection between `cusp-core',  `too-big-to-fail' and abundance matching relation was highlighted in \citet{brook15}: the authors used the dynamical mass at half-light radius for Local Group's dwarf galaxies to show that, when fitted by a mass-dependent cored profiles \citep{arianna}, the kinematic of galaxies with $M_{\star}>10^6 M_{\odot}$ is compatible with haloes more massive than $M_{\rm halo} \simeq 10^{10} M _{\odot}$, alleviating the `too-big-to-fail' problem and providing a $M_{\star} - M_{\rm halo}$ relation in line with abundance matching predictions.

Since the core creation can be explained with SN feedback (e.g. \citealt{penarrubia}, \citealt{arianna}, \citealt{maxwell15}, \citealt{onorbe15}, \citealt{chan15}, \citealt{tollet16}), here we  quantify its capability to modify the DM halo profiles at early times, by $z=2$, for our sample of dwarfs, by providing an observationally-based accurate determination of the total mass in stars formed in stars by $z=2$.

This paper is organized as follows: In Sect.~\ref{sec:methodology} we detail the data-sets and methodology used to correct the SFHs from the incomplete spatial sampling and obtain the mass formed into stars up to $z=2$. In Sect.~\ref{sec:mass} we compare this derived quantity with present-day observables such as the stellar and dynamical mass as a function of the \citet{gallart} dichotomy. In Sect~\ref{sec:feedback} we calculate the amount of SN feedback energy produced at early times (by $z=2$) and study its capability to remove the gaseous component and to change the DM density profiles of our dwarfs. We summarize our results and conclusions in Sect. \ref{sec:conclusions}. In Appendix~\ref{sec:lower} we investigate the effect of allowing for the feedback energy to be all injected at even earlier times ($z=6$), accounting for the expected lower DM halo masses; we compare our results to simulations and theoretical work in the literature and briefly make considerations on gas expulsion and core creation in fainter systems than those considered in the main text, such as ultra-faint dwarf galaxies.

\section{Methodology and data-sets}
\label{sec:methodology}
One of our goals is to compare the amount of stellar mass formed at ancient times in LG dwarf galaxies, in order to identify possible differences in the early properties of {\it slow} and {\it fast} dwarfs. Our sample consists of 16 LG `classical' dwarf galaxies\footnote{Even though the distinction might be somewhat artificial, here we mantain the common nomenclature of `classical' dwarf galaxies and `ultra faint' dwarf galaxies (UFDs) to refer broadly to Local Group dwarf galaxies whose existence was know prior and posterior to the advent of SDSS, respectively.}. Fifteen of them were drawn from the sample used in \citet{gallart}, from which we excluded the LMC, SMC and IC1613, since the spatial properties of their most ancient stellar component are still largely undetermined. We have also added the Leo~II dSph due to the recent availability of wide-area photometric catalogues (Stetson, private communication) and SFH (Monelli, private communication). A list of galaxies and their properties\footnote{We note that there are clear evidences that And~II has experienced a relatively high mass-ratio merger (at least 1:10, see \citealt{amorisco14}); this might have affected the observed properties of the stellar component of this system, such as its half-light ratio, and placed it out of dynamical equilibrium; at the same time, its observed SFH might be the mix of the SFHs from the two merging systems.} is given in Table~1.

An important aspect to take into account when determining the amount of stellar mass formed at a given redshift is the existence of negative age gradients in several LG dwarf galaxies (e.g. \citealt{bat6}, \citealt{bat12}, \citealt{bat12b}, \citealt{hi13}): for these small galaxies in general the old stars show a more spatially extended distribution than the younger ones. Since not all the LG dwarf galaxies benefit from SFHs derived from deep CMDs covering a large portion of the galaxy's stellar component, not correcting for the missing spatial coverage could result in underestimations of the stellar mass from SFH integration, in particular for the old, most spatially extended stellar populations.  

It is possible to calculate the correction factor due to the missing spatial coverage by knowing the surface density profile and the structural parameters (ellipticity, position angle) of the spatial distribution of $>$10 Gyr old stars in each dwarf galaxy and the footprint of the data-sets from which SFHs were derived. An accurate determination of the surface density profile and the structural parameters requires photometric data with a wide-area coverage. When available, we adopt the literature values derived from the radial distribution of stellar mass at lookback-times $>$10 Gyr ago from SFH determinations from very deep and spatially extended photometric data.  In lack of such estimates, a suitable alternative for our goals is to use the horizontal branch (HB) as a tracer of $>$10 Gyr old stars, as supported by stellar evolutionary models. The HB is also about 3 magnitude brighter than the oMSTO, therefore very wide-area photometric data reaching down to below the HB level are more much easily encountered in the literature/archives.

\subsection{Spatial distribution of ancient stars} \label{sec:spatial}
Tab.~\ref{tab:t1} lists the ellipticity, position angle $\theta$\footnote{Defined as the angle of the galaxy projected semi-major axis from North to East.} and scale length of the exponential profile, $R_s^{\rm old}$, that we adopted for the $>$10 Gyr old stellar component of each dwarf galaxy, together with the corresponding sources. As detailed below, the complete set of estimates was not available in the literature for all galaxies.

For And~II, there was no estimate of the best-fitting exponential surface density profiles. Therefore we use the S\'ersic profile of index $n$ = 0.3 by \citet{mc7}. 

For Draco, Ursa Minor, Carina, Leo~I and Leo~II, we have estimated ourselves the best fitting exponential surface density profile of the HB stars, and the corresponding structural parameters. To this aim we use photometric catalogues of point-sources derived from archive data: CFHT/MegaCam for Draco and Ursa Minor (Irwin, private communication), CTIO/MOSAIC~II for Carina \citep{bat12b}; compilations of data from different instruments for Leo I \citep{stetson14} and Leo II (Stetson, private communication). We then isolate HB stars using a simple selection in magnitude and color over the CMD\footnote{Incompleteness due to crowding or different depths between pointings is not an issue given the low surface brightness of the galaxies we are examining and the relatively bright apparent magnitude of the HB relative to the depth of the photometric data-sets.}. A detailed explanation of the methodology for fitting the structural parameters and best-fitting surface (number) density profile can be found in \citet{luis}. Here, it suffices to say that the analysis is performed by applying Bayesian MCMC methods directly to the stars' position, following the formalism described in the appendix of \citet{richardson}; we apply the MCMC Hammer \citep{foreman}, a Python implementation of the Affine Invariant MCMC Ensemble sampler \citep{goodman}. The dwarf galaxy's surface density profile is assumed to be an exponential profile; a constant term is added in order to account for contamination by fore/background sources. There are 7 free parameters in the fit: the central surface density ($\sigma_0$), exponential scale length ($R_s$), central coordinates ($\alpha_0$, $\delta_0$), position angle ($\theta$) and ellipticity ($\epsilon$) of the dwarf galaxy's stellar component, and the surface density of contaminants ($\sigma_c$).

We compared our results with the spatial distribution parameters from \citet{mc12} for the whole stellar population, finding always consistency with our results and the known presence or absence of population gradients in these systems \citep[e.g.][]{bat12b, jin}. 

For And~XVI and Aquarius there are no wide-area photometric catalogues reaching down to below the horizontal branch that we could access. Hence, for these two galaxies we use the half-light radius for the whole stellar component from \citet{mc12}, coupled with the transformation between scale length and half-light radius by \citet{wolf}. This gives a lower limit on the scale length of the old stars due to the possible presence of negative age gradients.

\subsection{Star formation histories}
The Star Formation Rates (SFRs) as a function of cosmic time (i.e., SFHs) derived from deep CMDs are used to calculate the amount of mass formed more than 10 Gyr ago within the region probed by these deep photometric data-sets. We summarize in Tab.~\ref{tab:t1} the sources of these data-sets and the type of each dwarf galaxy according to the classification proposed by \citet{gallart}.

For the Fornax dSph, since the SFH by \citet{pino13} comes from a very deep VLT/FORS photometric data-set but with a tiny spatial coverage, we carry out our analysis also with the SFH from \citet{boer12b}, whose CTIO/Mosaic~II data-set is not as deep, but has a spatial sampling that is almost complete.

\subsection{Correction for the incomplete spatial sampling}
We create mock galaxies following the surface density profile and structural parameters of the old stellar population (see Sect.~\ref{sec:spatial} and Tab.~\ref{tab:t1}). The procedure is based on the generation of 2-dimensional random arrays following the desired spatial distribution (see also Appendix B of \citealt{luis}).

We reproduce the spatial coverage of the observations from where the SFHs were derived, taking from the literature the size, shape, orientation and deviation with respect to the dwarf central coordinates of the footprint of each photometric data-set. Then, we define the \textit{coverage} as the percentage of mock stars from the generated galaxies that fall into the limits of the observed footprint.  
We list in Tab.~\ref{tab:t1} the obtained coverage percentages, for which one can see a wide range of values: from very well covered galaxies, such as  Sculptor, Tucana or Carina, to observations that are missing the majority of the old stellar component, like the case of And~II or Fornax for the \citealt{pino13} data-set. It can be appreciated that the correction is non-negligible in several cases. The uncertainties in the coverage percentage were obtained by propagating the error in the scale length of the old stellar population $R_s^{\rm old}$ and neglecting the errors in the ellipticity and the position angle, since the coverage determination is strongly dominated by $R_s^{\rm old}$.

As final step, we integrate the SFHs from the beginning of star formation to 10 Gyr ago ($z \simeq 2$). The resulting mass is then divided by the corresponding coverage percentage, and this yields the corrected mass formed into stars up to $z=2$, $M_{\star, z>2}$. The uncertainties on $M_{\star, z>2}$ are calculated considering the intrinsic error of the SFRs and the error introduced by our procedure (i.e. the error in the coverage).

\begin{table*}

\caption{Structural parameters, derived parameters and properties of the 16 analyzed dwarf galaxies: ellipticity, position angle and 2D scale length adopted for the old stellar component ($\epsilon$, $\theta$, $R_s^{\mathrm{old}}$ ), classification based on the SFH, coverage percentage, mass formed into stars up to $z=2$ ($M_{\star, z>2}$), present-day luminosity in V band ($L_V$), mass-to-light ratio for the $V$ band ($M/L$), heliocentric distance ($d$), line-of-sight velocity dispersion ($\sigma_{v}$) and 2D half-light radius of the overall stellar component ($R_{1/2}$). }
\label{tab:t1}

\begin{tabular}{lccccccccccc} 
	\hline 
Galaxy & $\epsilon$ & $\theta$ & $R_s^{\mathrm{old}}$   & SFH & Coverage   &  $M_{\star, z>2}$  & $L_V$ & $M/L$ & $d$  & $\sigma_{v}$  & $R_{1/2}$ \\
& & (\degree) & (arcmin) & & (\%) & $(10^6 M_{\odot}$) & $(10^6 L_{\odot}$) &  & (kpc) & (km s$^{-1}$) & (pc)\\
		\hline

Cetus & 0.33$^{(2)}$ & 63$^{(2)}$ &  0.98$\pm$0.02$^{(5)}$  & fast$^{(12)}$  & 28.1$_{-0.5}^{+0.4}$ & 5.99$_{-0.34}^{+0.34}$ & 2.60$^{(2)}$ & 1.6$^{(3)}$ & 790$^{(32)}$ & 17$\pm$2$^{(2)}$ & 703$\pm$31$^{(2)}$\\ 

Tucana & 0.48$^{(2)}$ & 97$^{(2)}$ &  0.54$\pm$0.03$^{(5)}$  & fast$^{(13)}$  & 92.7$_{-1.3}^{+1.2}$ & 2.39$_{-0.14}^{+0.18}$ & 0.56$^{(2)}$ & 1.6$^{(3)}$ & 899$^{(33)}$ & 15.8$_{-3.1}^{+4.1}$$^{(2)}$ & 284$\pm$54$^{(2)}$\\ 

LGS-3 & 0.20$^{(2)}$ & 0$^{(2)}$ &  0.87$\pm$0.09$^{(5)}$  & fast$^{(14)}$  & 59.5$_{-4.9}^{+4.1}$ & 1.57$_{-0.24}^{+0.32}$ & 0.96$^{(2)}$ & 1.0$^{(3)}$ & 650$^{(31)}$ & 7.9$_{-2.9}^{+5.3}$$^{(2)}$ & 470$\pm$47$^{(2)}$\\

Leo A & 0.40$^{(2)}$ & 114$^{(2)}$ &  1.61$\pm$0.68$^{(8)}$  & slow$^{(15)}$  & 42.0$_{-15.0}^{+25.0}$ & 0.66$_{-0.44}^{+0.91}$ & 6.00$^{(2)}$ & 0.5$^{(3)}$ & 798$^{(2)}$ & 9.3$\pm$1.3$^{(4)}$ & 354$\pm$19$^{(4)}$\\ 

And II & 0.20$^{(2)}$ & 34$^{(2)}$ &  10.0$\pm$8.5$^{(11)}$  & fast$^{(16)}$  & 5.0$_{-3.5}^{+58.0}$ & 54$_{-50}^{+162}$ & 7.60$^{(2)}$ & 1.0$^{(3)}$ & 652$^{(2)}$ & 7.3$\pm$0.8$^{(2)}$ & 1176$\pm$50$^{(2)}$\\ 

And XVI & 0.00$^{(2)}$ & 0$^{(2)}$ &  0.53$\pm$0.03$^{(2)}$  & slow$^{(17)}$  & 71.4$_{-2.1}^{+2.2}$ & 0.21$_{-0.07}^{+0.13}$ & 0.41$^{(2)}$ & 1.2$^{(3)}$ & 525$^{(2)}$ & 3.8$\pm$2.9$^{(4)}$ & 136$\pm$15$^{(2)}$\\

Draco & 0.22$^{(1)}$ & 82$^{(1)}$  &  5.41$_{-0.29}^{+0.31}$$^{(1)}$  & fast$^{(18)}$  & 90.8$_{-1.9}^{+1.3}$ & 0.56$_{-0.15}^{+0.15}$ & 0.18$^{(29)}$ & 1.8$^{(3)}$ & 76$^{(2)}$ & 9.1$\pm$1.2$^{(2)}$ & 221$\pm$19$^{(2)}$\\

UMi & 0.52$^{(1)}$ & 48$^{(1)}$ &  10.9$_{-0.6}^{+0.7}$$^{(1)}$  & fast$^{(19)}$  & 74.7$_{-2.4}^{+1.9}$ & 0.88$_{-0.11}^{+0.13}$ & 0.29$^{(2)}$ & 1.9$^{(3)}$ & 76$^{(2)}$ & 9.5$\pm$1.1$^{(2)}$ & 411$\pm$31$^{(2)}$\\

Sculptor & 0.32$^{(2)}$ & 99$^{(2)}$ &  9.0$\pm$0.3$^{(7)}$  & fast$^{(20)}$  & 99.3$_{-0.2}^{+0.2}$ & 6.78$_{-0.60}^{+0.60}$ & 2.30$^{(2)}$ & 1.7$^{(3)}$ & 86$^{(2)}$ & 9.2$\pm$1.4$^{(2)}$ & 283$\pm$45$^{(2)}$\\

Carina & 0.28$^{(1)}$ & 67$^{(1)}$ &  7.78$_{-1.15}^{+1.39}$$^{(1)}$  & slow$^{(21)}$  & 92.7$_{-4.5}^{+2.6}$ & 0.31$_{-0.14}^{+0.16}$ & 0.38$^{(2)}$ & 1.0$^{(3)}$ & 105$^{(2)}$ & 6.6$\pm$1.2$^{(2)}$ & 250$\pm$39$^{(2)}$\\

Phoenix & 0.40$^{(2)}$ & 5$^{(2)}$ &  1.56$\pm$0.05$^{(6)}$  & fast$^{(22)}$  & 37.0$_{-1.2}^{+0.9}$ & 2.40$_{-0.27}^{+0.25}$ & 0.77$^{(2)}$ & 1.8$^{(3)}$ & 415$^{(2)}$ & 9.3$\pm$0.7$^{(28)}$ & 274$\pm$8$^{(8)}$\\

Leo I & 0.35$^{(1)}$ & 77$^{(1)}$ &  2.35$_{-0.09}^{+0.09}$$^{(1)}$  & slow$^{(23)}$  & 13.1$_{-0.9}^{+0.9}$ & 0.39$_{-0.25}^{+0.28}$ & 5.50$^{(2)}$ & 0.9$^{(3)}$ & 254$^{(2)}$ & 9.2$\pm$1.4$^{(2)}$ & 251$\pm$27$^{(2)}$\\

Leo II & 0.13$^{(2)}$ & 12$^{(2)}$ &  1.99$_{-0.14}^{+0.15}$$^{(1)}$  & slow$^{(24)}$  & 13.9$_{-1.4}^{+2.1}$ & 0.77$_{-0.25}^{+0.28}$ & 0.74$^{(2)}$ & 1.6$^{(3)}$ & 233$^{(2)}$ & 6.6$\pm$0.7$^{(2)}$ & 176$\pm$42$^{(2)}$\\

Aquarius & 0.50$^{(2)}$ & 99$^{(2)}$ &  0.88$\pm$0.02$^{(2)}$  & slow$^{(25)}$  & 77.6$_{-1.5}^{+1.6}$ & 0.67$_{-0.31}^{+0.30}$ & 1.60$^{(2)}$ & 0.9$^{(3)}$ & 1072$^{(2)}$ & 7.9$_{-1.6}^{+1.9}$$^{(4)}$ & 458$\pm$21$^{(2)}$\\ 

Sextans & 0.27$^{(2)}$ & 52$^{(2)}$ &  12.7$_{-0.4}^{+0.4}$$^{(9)}$  & fast$^{(26)}$  & 51.4$_{-2.0}^{+1.9}$ & 3.21$_{-0.94}^{+1.02}$ & 0.44$^{(2)}$ & 1.6$^{(3)}$ & 86$^{(2)}$ & 7.9$\pm$1.3$^{(2)}$ & 695$\pm$44$^{(2)}$\\

Fornax & 0.30$^{(2)}$ & 41$^{(2)}$ &  13.7$\pm$0.2$^{(10)}$  & slow$^{(27)}$  & 2.2$_{-0.1}^{+0.1}$ & 18.9$_{-3.9}^{+4.8}$ & 20.0$^{(2)}$ & 1.2$^{(3)}$ & 147$^{(2)}$ & 11.7$\pm$0.9$^{(2)}$ & 710$\pm$77$^{(2)}$\\

 &  &  &    & slow$^{(30)}$  & 88.2$_{-0.5}^{+0.5}$ & 3.8$_{-0.8}^{+0.8}$ & & &  &  & \\
\hline
	\end{tabular}
\\

\begin{tabular}{llll}
$^{(1)}$This work (MCMC Hammer) & $^{(2)}$\citet{mc12} & $^{(3)}$\citet{woo} &
$^{(4)}$\citet{kirby} \\

$^{(5)}$\citet{hi13}  & 
$^{(6)}$\citet{bat12}   &$^{(7)}$\citet{thesis}  & $^{(8)}$Hidalgo private communication \\

$^{(9)}$\citet{luis} &  $^{(10)}$\citet{bat6} &   $^{(11)}$\citet{mc7} &  $^{(12)}$\citet{mateo10a} \\

$^{(13)}$\citet{mateo10b} &  $^{(14)}$\citet{hi11} & $^{(15)}$\citet{cole7} &$^{(16)}$\citet{skil16} \\

$^{(17)}$\citet{mateo16} &  $^{(18)}$\citet{aparicio} & $^{(19)}$\citet{carrera02} &   $^{(20)}$\citet{boer12} \\

 $^{(21)}$\citet{boer14} & $^{(22)}$\citet{hi9}
 & $^{(23)}$\citet{gallart99}  & $^{(24)}$Monelli private communication \\

$^{(25)}$\citet{cole14} &
$^{(26)}$\citet{lee9} &
$^{(27)}$\citet{pino13} & $^{(28)}$\citet{kacharov}   \\

$^{(29)}$\citet{ih95} & $^{(30)}$\citet{boer12b}  & $^{(31)}$\citet{bernard} & $^{(32)}$\citet{castellani} \\

 $^{(33)}$\citet{sarajedini}

\end{tabular}

\end{table*}

\section{MASS FORMED INTO STARS UP TO $Z=2$} \label{sec:mass}
In this section we compare our derived value of $M_{\star, z>2}$ with present-day observables such as the stellar mass (Fig.~\ref{fig:mass}) and the dynamical mass within the half-light radius (Fig.~\ref{fig:mdyn}). We study the behavior of these relations as a function of the {\it fast} and {\it slow} classification.


Due to the existence of age gradients, the  relative fraction of old stellar population could have been underestimated in galaxies classified as {\it slow} in \citet{gallart} and whose SFH was determined from data covering a small fraction of the main body. Thus, we first redefined a classification criterion based on our absolute quantities and checked whether our determination of $M_{\star, z>2}$ would result in a different classification than the one proposed by \citet{gallart}. 

Ideally, one would want to use SFHs corrected for the missing spatial coverage for the whole lifetime of the galaxy. However, this would imply knowledge of the dependence of the scale length of the spatial distribution of the stellar component as a function of age, which is not available. Instead, we calculate how much the old stellar component is contributing to the total present-day stellar mass.  To this end, we use \textsc{iac-star} \citep{iacstar} in order to calculate the fraction of $M_{\star, z>2}$ that remains alive to present-day, $M_{\star, z>2}^{\rm alive}$. By assuming a constant SFR between $t \simeq 13.5$ and $t = 10$ Gyr in lookback time and a typical metal-poor population, we obtain that $\sim$40\% of the mass formed in stars before 10 Gyr ago is still alive today ($\sim$60\% when including remnants). Therefore, we adopt a 50\% factor to obtain $M_{\star, z>2}^{\rm alive}$. We checked that this number does not change significally if the shape of the SFH at old times is different from constant, e.g. assuming it all concentrated within the first $\sim$1 Gyr. Then, we compare the obtained $M_{\star, z>2}^{\rm alive}$ with the present-day total stellar mass $M_{\star}^{T}$, that is derived using the luminosities from Tab.~\ref{tab:t1} and the stellar component mass-to-light ratios by \citet{woo}\footnote{\citet{woo} do not list $M/L$ values for Cetus and And XVI. For those galaxies, we adopt the same $M/L$ than for Tucana and Fornax (respectively) based on the similarity of their SFHs.}. We define a dwarf galaxy as {\it fast} if $M_{\star, z>2}^{\rm alive}$ > ($M_{\star}^{T}$ - $M_{\star, z>2}^{\rm alive}$), i.e. when more than 50\% of its current mass in stars was formed before $z = 2$, or equivalently, when $M_{\star, z>2} > M_{\star}^{T}$ . According to this criterion, all the analyzed galaxies remain of the same type as in \citet{gallart}. We note that Sextans falls off the permitted region in amount of ancient stars that could have formed given the present-day luminosity, but considering the errorbars, the discrepancy is not statistically significant.

\subsection{Relation to the present-day stellar mass}
\begin{figure*}
\includegraphics[width = 0.7\textwidth]{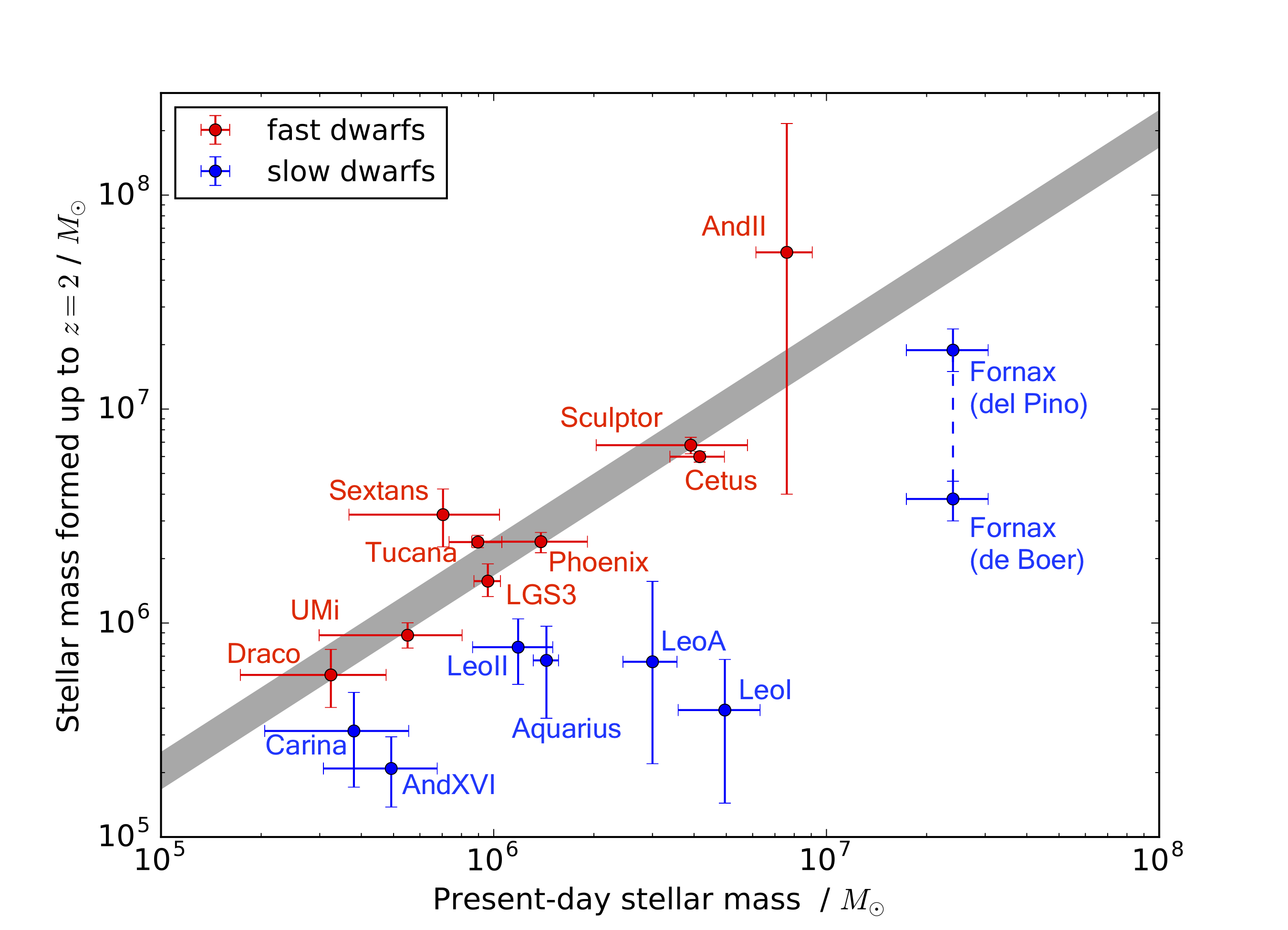}

\caption{Stellar mass formed up to $z = 2$ ($M_{\star, z>2}$) as a function of the mass in stars still alive at the present-day ($M_{\star}^{T}$). The red and blue points represent respectively the {\it fast} and {\it slow} dwarfs according to the classification by \citet{gallart}. The x-axis errors are obtained propagating the uncertainties on $M_V$ from \citet{mc12}. The y-axis errors are calculated considering the intrinsic error of the SFRs and the error introduced by our procedure (i.e. the error in the coverage). The grey band indicates the maximum $M_{\star,z>2}$ allowed as a function of present-day stellar mass, with the limits corresponding to this calculation done with remnants and without remnants once the effect of stellar evolution and death are taken into account. }
\label{fig:mass}
\end{figure*}
In the work of \citet{gallart} the life-time evolution of {\it fast} and {\it slow} dwarfs was compared in relative terms, using normalized SFHs. Figure~\ref{fig:mass} shows that, when comparing galaxies at a similar present-day stellar mass, dwarfs classified as \textit{fast} appear to have formed more stellar mass at ancient times than \textit{slow} types, also in absolute terms. It could be argued that the above result may be intrinsic to the definition of {\it fast} and {\it slow} types; however, here we showed that {\it the result holds also when correcting the amount of stellar mass formed at ancient times for the missing spatial coverage}, which in several cases implied a significant correction. Furthermore, the trend of {\it fast} dwarfs being more massive at $z=2$ than {\it slow} types appears to hold over about two orders of magnitude in current stellar mass. Information on the SFH of a system like WLM,  of comparable stellar mass to Fornax but star forming and gas-rich, would help in testing whether the trend continues at stellar masses above $10^7 M_{\odot}$.

The values of $M_{\star, z>2}$ in And~II and Fornax are the most uncertain, due to the very large error associated to $R_s^{\rm old}$ for the former and due to the different SFH determinations in the literature for the latter. Nonetheless, it is very likely that Fornax had the largest baryonic mass of the galaxies in the sample since early on, as it would be suggested by the fact that it contains five ancient globular clusters, while no GCs have been detected in And~II, Sculptor and Cetus. 

An additional interesting information provided by Fig.~\ref{fig:mass} is that all but one of the {\it slow} dwarfs in the sample formed a similarly low amount of stellar mass at early times ($2 \times 10^5 < $ M$_{\odot} < 8 \times 10^5$), independently of their current stellar mass. Since these galaxies have all experienced star formation more recently than $z=2$ and still contain gas at present, this may indicate that the supernovae feedback associated with early stellar masses below $10^6 M_{\odot}$ would be insufficient to induce a significant removal of gas and therefore would not result in an early quenching of the star formation. 

On the other hand, the Milky Way satellites Draco and Ursa Minor have also formed similarly low stellar masses at ancient times but have had their star formation stopped by $z=2$;
we speculate then that star formation in these latter systems was not stopped by internal feedback alone (see also Sect.~\ref{sec:gas}), but rather by other effects, such as ram-pressure stripping and/or reionization. 

For example, \citet{gallart} proposed that strong internal feedback and reionization may couple to induce important gas loss at early times leading to early quenching in {\it fast} dwarfs, while this coupling may not occur in slow dwarfs, in which the onset of star formation would be delayed and take place only when the dark matter halo has grown massive enough to allow the gas to cool and form stars after reionization. A similar dichotomy in SFHs was noticed in work by \citet{benitez15}, which makes use of the CLUES simulations (\citealt{gottloeber}, \citealt{yepes}). The dwarf galaxies with predominantly old SFHs are those inhabiting DM haloes that collapse early and where re-ionization coupled with internal feedback drives the low-density gas out of the virial radius, preventing further re-accretion: therefore star formation continues only until the gas that had cooled down before re-ionization is eventually consumed. On the other hand, the dwarf galaxies with a predominantly young stellar population are inhabiting haloes that collapsed late
and were thus unable to start forming stars in significant numbers until well after reionization. Several sets of simulations do lend support to the fact that the heating effect of internal feedback is enhanced by the UV background due to reionization (e.g. \citealt{sawala10}). Note that the effects of reionization could be  spatially dependent, being stronger for those galaxies that were born closer to a growing source of reionization like the MW or M31 and affecting in a higher amount satellites like Draco and Ursa Minor (\citealt{miralda00}, \citealt{weinmann07}, \citealt{spitler}, \citealt{ocvirk}, \citealt{dixon}).

If Draco and Ursa Minor would have not been stripped of their gaseous component and been allowed to continue forming stars till present day, they might have turned out to be much more luminous, as their ``slower" Leo~A, Leo~I counterparts. Therefore, when exploring the hypothesis that passively evolving satellites of the MW and M31 shared similar ancestors as the gas-rich, star forming isolated LG dwarfs, depending on the infall redshift of the satellite, one might have to compare systems of rather different present-day luminosity (see also \citealt{mistani} for similar conclusions between cluster and field dwarf galaxies).

\subsection{Relation to the dynamical mass} \label{sec:dm}

Since at similar present-day stellar mass dwarf galaxies with a fast SFH appear to have formed a larger amount of stellar mass at early times than slow dwarfs, it is interesting to explore whether there are signs they could have assembled also more dark matter (DM) at early times. 

To this aim, we consider the dynamical mass within the half-light radius ($M_{\rm dyn}$), which can be derived very accurately in pressure-supported spherical systems for which only l.o.s. velocities are available, such as the gas-poor dSphs, provided a few conditions are met \citep[see][]{walker9, wolf}. Many of the gas-rich LG dwarfs show little sign of rotation in their stellar component within their projected half light radii \citep{leaman2012,wheeler}, therefore the same method has been applied to these systems too (e.g. \citealt{kirby}). 
\begin{figure*}
\includegraphics[width = \textwidth]{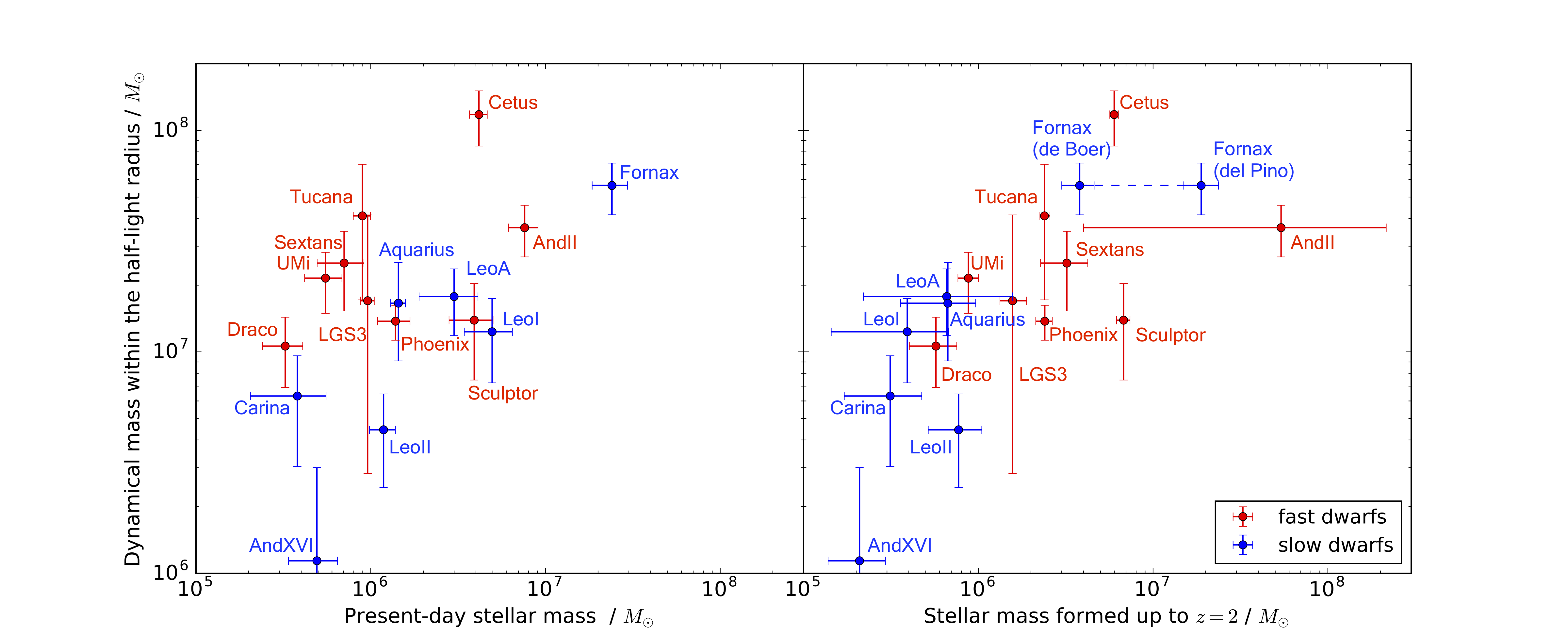}
\caption{Dynamical mass within the half-light radius ($M_{\rm dyn}$) as a function of the present day stellar mass ($M_{\star}^T$, left panel) and stellar mass formed up to $z = 2$ ($M_{\star, z>2}$), right panel). The red and blue points represent respectively the {\it fast} and {\it slow} dwarfs according to the classification by \citet{gallart}.}
\label{fig:mdyn}
\end{figure*}

Here we use the formula by \citet{walker9}:
\begin{equation}
M_{\rm dyn } [M_{\odot}] = 580 R_{1/2} \sigma_{v}^2 
\end{equation}
where $R_{1/2}$ is the two-dimensional half-light radius in pc, $\sigma_v$ is the l.o.s. velocity dispersion in km s$^{-1}$, for which we use the values referring to the overall stellar component (see Tab.~\ref{tab:t1}). Since the Local Group dwarf galaxies we are analyzing are consistent with being dark matter dominated at all radii (see recent  reviews by \citealt{bat13}, \citealt{walker} and references therein, although see \citealt{bat15} and \citealt{diako} for the case of Fornax), $M_{\rm dyn }$ is essentially due to the mass of the DM halo within the half-light radius\footnote{We checked that the stellar mass within the half-light radius is negligible with respect to the total $M_{\rm dyn}$ in our dwarfs.}. 
The effect of tidal stripping on to the DM haloes of satellite dwarf galaxies is to decrease the intrinsic $\sigma_v$ (e.g. \citealt{read06}, \citealt{penarrubia08}, \citealt{lokas11}, \citealt{kazantzidis}) and to a less extent $R_{1/2}$ \citep{penarrubia08}; therefore the $M_{\rm dyn}$ value for satellite dwarf galaxies should be seen as a lower limit to the $M_{\rm dyn }$ before infall.  

Figure~\ref{fig:mdyn} shows that there are hints of a correlation between the present-day stellar mass and $M_{\rm dyn }$ (left panel, see also McConnachie 2012); this correlation becomes better defined when considering the stellar mass formed at early times $M_{\star, z>2}$ (right panel), rather than the present-day stellar mass. This is confirmed by a Pearson test, which yields a correlation coefficient equal to 0.6 for log($M_{\rm dyn })$ vs log($M_{\star}^{T}$) and 0.7 for log($M_{\rm dyn }$) vs log ($M_{\star, z>2}$).

We speculate that the relation between $M_{\rm dyn}$ and $M_{\star, z>2}$ might be linking the DM and the baryonic content of these dwarf galaxies at early times, with $M_{\star, z>2}$ being a better tracer of the initial baryonic content than the present-day stellar mass. This is probably because, as discussed above, for systems which have had their evolution affected by external mechanisms, the present-day stellar mass is likely not representative of the stellar mass the system would have had if allowed to evolve in isolation. 

We find that, in general, {\it fast} dwarfs have a larger dark matter content within the half-light radius with respect to {\it slow} dwarfs: excluding And~II, which seems to be an outlier in terms of $R_{1/2}$, the median $M_{\rm dyn}$  is $1.9 \times 10^7 M_{\odot}$ for the {\it fast} ($2.2 \times 10^7 M_{\odot}$ if we do not exclude And~II) and $1.2 \times 10^7 M_{\odot}$ for the {\it slow} dwarfs, respectively. This could, however, be a consequence of the larger $R_{1/2}$ found in {\it fast} dwarfs for the galaxies in our sample: the ratio between the median $R_{1/2}$ of both kind of dwarfs is $\sim$1.4 excluding And II ($\sim$1.6 with And~II), while the ratio between median dynamical masses is $\sim$1.5-1.6 ($\sim$1.8 with And~II). Given the relatively small size of our sample of galaxies, we cannot exclude that small number statistics might be affecting the result. Nonetheless we note that: 1) the DM mass for the {\it slow} dwarfs which are gas-rich could be lower than estimated here, since we have neglected the contribution of the gas to $M_{\rm dyn}$; 2) the DM mass (and hence $M_{\rm dyn}$) for the dwarfs which are satellite galaxies, that is in general {\it fast} dwarfs, is expected to have been larger before infall, depending on the amount of tidal stripping undergone. These effects should, on average, go in the direction of emphasizing the difference in DM mass between {\it slow} and {\it fast} dwarfs.

\section{Supernova feedback} \label{sec:feedback}

We use our determination of the mass in stars formed up to $z = 2$ to provide observationally motived estimates of the amount of stellar feedback from supernovae (SNe) explosions and comment on the possible effect that this might have had on the early evolution of the baryonic and dark matter component of Local Group dwarf galaxies. Specifically, we focus on the questions of whether the energy injected by SN is able to remove the gas and therefore halt star formation at early times in the {\it fast} dwarfs (Sect.~\ref{sec:gas}) and/or to transform an initially cuspy dark matter halo into a cored one (Sect.~\ref{sec:cores}). 

We concentrate on SNe~II because 3D hydrodynamical simulations show that the influence of the SNe~Ia explosions on the general hydrodynamical behaviour of the ISM is not very important, due to the small percentage ($\sim$3\%) of SNe~Ia events during a cycle of SNe~II explosions (e.g. \citealt{marcolini}). 

 \subsection{Derivation of SNae feedback energy and competing gravitational potential}
 
In order to calculate the expected energy budget from SNe~II explosions ($E_{\rm SN, z=2}$), we assume that stars with masses $M \gtrsim 6.5 M_{\odot}$ evolve into SNe~II and calculate the expected numbers integrating the IMF by \citet{kroupa}, given our estimates of $M_{\star, z>2}$. Considering a typical kinetic energy of $10^{51}$ erg for these kind of events \citep{utrobin}, we obtain the total energy injected by SNe II in the environment up to $z \simeq 2$. We list in Tab.~\ref{tab:t2} the values of $E_{\rm SN, z=2}$. The choice of a SN~II cut-off mass of $6.5 M_{\odot}$ was motivated by theoretical works \citep[see e.g.][]{cassisi1993, mateo10b} predicting a lower value at the low metallicity of stellar populations in dwarf galaxies, with respect to the standard $8\pm 1 M_{\odot}$ cut-off observed at solar, or higher, metallicity \citep[see e.g. review by][]{smartt2009}. We checked that the effect of choosing $M \gtrsim 8 M_{\odot}$ as limit in mass for the integration does not change significantly the result: in that case, the feedback energy obtained is a $\sim$75\% of the $E_{\rm SN, z=2}$ in the $M \gtrsim 6.5 M_{\odot}$ case (see below for consequences on the results). 

A key factor that strongly affects the analysis is the choice of what fraction of the energy produced by SN~II couples to the gas and to the dark matter; this is parametrized through the so-called efficiency ($\epsilon_{\rm SN}$ and $\epsilon_{\rm DM}$). The value of this parameter is very uncertain, but as discussed for example in \citet{penarrubia}, typical values are likely not to exceed $\epsilon_{\rm SN}$ = 0.4 \citep[see e.g.][whose simulations also include heating from the cosmic UV background]{governato} and might be as low as a few percent ($\epsilon_{\rm SN} = 0.01$, e.g. \citealt{kellermann}; in \citealt{revaz} a value of $\epsilon_{\rm SN}$ = 0.05 best describes the metallicity-luminosity relation of MW dSphs). 3D hydrodynamical simulations of the chemical and dynamical evolution of the ISM in dwarf galaxies show that, in a picture where star formation proceeds in short bursts of 60 Myr, even if the energy released by SNe II in a {\it single} burst is about 1.3 times larger than the gas binding energy, no galactic wind develops due to the massive DM halo and the large effectiveness of the radiative losses \citep{marcolini}; also in this case the SN efficiency appears to be $\sim$0.05. The range of efficiencies from observationally motivated works is compatible with the theoretical results, e.g. \citet{mcquinn} calculate an average wind efficiency of 0.16 for a burst timescale of 25 Myr.

Another crucial ingredient to quantify the capability of the SNe feedback energy of removing the gas and/or modifying the dark matter halo density profile by $z = 2$ is the competing effect of the gravitational potential $W$ of the dwarf galaxy. As discussed above, Local Group dwarf galaxies are typically found to be DM dominated at all radii; therefore we neglect the stellar component as a contributor to the gravitational potential (note that adding it would go in the direction of making gas removal or creation of cores in the DM halo more difficult).  
We assume a standard $\Lambda$CDM cosmology with $\Omega_{\mathrm{m}}$ = 0.32, $\Omega_\Lambda$ = 0.68 and $H_0$ = 67 km s$^{-1}$ Mpc$^{-1}$ (e.g. \citealt{planck}).

The gravitational potential is defined as:
\begin{equation}
W = -4 \pi G \int_0^{R_{\mathrm{vir}}} \rho (r) M (r) r dr
\label{weq}
\end{equation}
where $\rho(r)$ is the DM halo density profile, $M(r) = \int_0^r \rho(r') 4 \pi r'^2 dr'$ and $R_{\mathrm{vir}}$ is the DM halo virial radius. 
The value of $W$ at $z \simeq 2$ is calculated as follows.

\noindent First, we obtain the DM halo mass at $z = 0$ from the present-day stellar mass using the AM relation by \citet{brook14} based on Local Group simulations:
\begin{equation}
\label{pl}
M_{\star} = \left(\frac{M_{\mathrm{h}}}{M_0 \times 10^6}\right)^{3.1}
\end{equation}
where $M_{\star}$ is the stellar mass, $M_{\rm h}$ is the halo mass and $M_0$ = 79.6 $M_{\odot}$. We also do the calculations  using  the AM relation from \citet{moster}:
\begin{equation}
\label{moster}
\frac{M_{\star}}{M_{\rm h}} = 2 N \left[\left(\frac{M_{\rm h}}{M_1}\right)^{-\beta}+\left(\frac{M_{\rm h}}{M_1}\right)^{\gamma}\right]^{-1}
\end{equation}
where $M_1$ = 11.59 $M_{\odot}$, $N$ = 0.0351, $\beta$ = 1.376 and $\gamma$ = 0.608. 

\noindent Second, we extrapolate the $z = 0$ DM halo mass to $z \simeq 2$ using the results by \citet{fakhouri} (see their figure 6), that for haloes in our regime of present-day mass ($\sim 10^{9-10} M_{\odot}$) estimates the virial mass $M_{\mathrm{vir}}$  at $z=2$ to be about 40-50\% of the current one (we have assumed $M_{\mathrm{vir}} \simeq M_{\mathrm{halo}}$).

\noindent Then, we obtain $R_{\mathrm{vir}} (z=2)$ from $M_{\mathrm{vir}} (z=2)$ following the formula\footnote{Here,  $\rho_c$ is the critical density of the Universe and $\Delta_{\mathrm{vir}}$ is the virial overdensity, that for a flat cosmology can be approximated by \citep{bryan}: $\Delta_{\mathrm{vir}}(z) \simeq \frac{18\pi^2 + 82x - 39x^2 }{\Omega_{\mathrm{m}}(z)}$,
with $x = \Omega_{\mathrm{m}}(z) - 1$, being $\Omega_{\mathrm{m}}(z)$ the normalized matter density, whose redshift evolution is related to the present-day matter density $\Omega_{\mathrm{m}}$ by the following formula: $\Omega_{\mathrm{m}}(z) =  \Omega_{\mathrm{m}} \frac{(1+z)^3}{1 - \Omega_{\mathrm{m}} + (1+z)^3 \Omega_{\mathrm{m}} }.$}:

\begin{equation}
M_{\mathrm{vir}} (z) = \frac{4\pi}{3} \Delta_{\mathrm{vir}} (z) \rho_c R_{\mathrm{vir}}^3(z)
\end{equation}

\noindent For the density profile $\rho(r)$ we assume a Navarro-Frenk-White (NFW) profile \citep{nfw}:
\begin{equation}
\rho_{\mathrm{NFW}} (r) = \frac{\rho_s}{(r/r_s) (1+r/r_s)^2}
\label{nfw}
\end{equation}
where to obtain the parameters $\rho_s$ and $r_s$ of the NFW profile it is useful to consider the concentration parameter $c_{\mathrm{vir}} \equiv R_{\mathrm{vir}}/r_s$. We calculate the concentration from $M_{\mathrm{vir}}(z=2)$ using the formula by \citet{dutton}:
\begin{equation}
\log_{10} c_{\mathrm{vir}} = a + b\log_{10}(M_{\mathrm{vir}}/[10^{12} h^{-1} M_{\odot}])
\label{cvir}
\end{equation}
with $a = 0.643$ and $b = -0.051$ for $z = 2$. Thus, $r_s$ is trivially obtained from $R_{\mathrm{vir}}$ and for $\rho_s$ by applying the relation in \citet{bullock}:
\begin{equation}
M_{\mathrm{vir}} = 4\pi \rho_s r_s^3 A(c_{\mathrm{vir}})
\end{equation}
where $A(c_{\mathrm{vir}})$ is defined as:
\begin{equation}
A(c_{\mathrm{vir}}) \equiv \log (1 + c_{\mathrm{vir}}) - \frac{c_{\mathrm{vir}}}{1+c_{\mathrm{vir}}}
\end{equation}

We decided not to use AM relations computed at $z=2$ in order to get our stellar to halo mass, because the relation is poorly constrained at such redshift for halo  masses below $10^{11}-10^{12}$ $M_{\odot}$. To check that our results do not depend strongly on the AM relation choice, we use both the \citet{brook14} AM relation (hereafter B14), which is based on Local Group simulations and observations, and the \citet{moster} AM relation (hereafter M13) extrapolated to low mass galaxies. We list in Tab. \ref{tab:t2} the virial masses at $z=2$ and gravitational potentials obtained for both AM relations. In Appendix~\ref{sec:lower} we comment on the effect of adopting the \citet{behroozi} AM relation, which is the one that differs the most from the two above.

Note that the use of AM relations for dwarf galaxies that are satellites results in an upper limit for their halo mass, as we derive the pre-infall halo masses, while the masses at z=0 could be much lower due the tidal stripping occurring after infall.

Simulations suggest that star formation proceeds in an oscillatory fashion, with 50-100 Myr long bursts, followed by similarly long quiescient periods, and that the impulsive heating due to this behaviour accumulates with time, eventually causing the transformation from a DM cusp into a core (e.g. \citealt{governato}, \citealt{teyssier}, \citealt{arianna}), depending on the energy balance. As we will show in the Appendix, the results from our energetics calculations are in very good agreement with those from hydrodynamical simulations of dwarf galaxy formation by \citet{read} when comparing the similar levels of energetics involved.

\subsection{Early gas removal} \label{sec:gas}

\begin{table*}
\caption{Feedback and gravitational potential properties of the dwarfs at $z=2$: budget of feedback energy (E$_{\rm SN}$), virial mass of the DM halo ($M_{\rm vir, z=2}$), gravitational potential of the DM halo ($W_{z=2}$) and minimum energy required to expel the gas ($\Delta W_{\rm gas/2}$) and form a core of $r_c$ = $R_{1/2}$ ($\Delta W_{\rm core}/2$), calculated for both AM relations by B14 and M13.}

\label{tab:t2}

\begin{tabular}{lccccccccccc}
\hline
Galaxy & E$_{\rm SN}$ ($10^{54}$ erg)  & \multicolumn{2}{c}{$M_{\rm vir, z=2}$ ($10^{9}$ $M_{\odot}$)} & \multicolumn{2}{c}{$W_{z=2}$ ($10^{54}$ erg)} & \multicolumn{2}{c}{$\Delta W_{\rm gas}/2$ ($10^{54}$ erg)} & \multicolumn{2}{c}{$\Delta W_{\rm core}/2$ ($10^{54}$ erg)}\\

 &  & B14 & M13 & B14 & M13 & B14 & M13 & B14 & M13 \\  \hline
Cetus &  83   & 4.9 & 4.3 & -60.8 & -49.8  & 5.1 & 4.1 & 5.6 & 4.7 \\
Tucana & 33 & 3.0 & 2.3 & -26.9 & -17.2 & 2.2 & 1.4 & 1.4 & 1.0 \\
LGS-3 & 22  & 3.0 & 2.3 & -27.8 & -18.0 & 2.3 & 1.5 & 
2.2 & 1.5\\
Leo A & 9 &   4.4 & 3.8 & -51.1 & -39.7 & 4.3 & 3.3 & 2.8 & 2.3\\
And II & 748  & 5.9 & 5.6 & -83.8 & -75.7 & 7.0 & 6.3 &  9.5 & 9.8 \\
And XVI & 3 & 2.5 & 1.8 & -19.5 & -11.2 & 1.6 & 0.9 & 0.6 & 0.4 \\
Draco & 8 & 2.1 & 1.5 & -15.5 & -8.4& 1.3 & 0.7 &  
0.7 & 0.4 \\
UMi & 12 & 2.6 & 1.9 & -20.7 & -12.2 & 1.7 & 1.0 & 
1.6 & 1.0\\
Sculptor & 94  & 4.8 & 4.2 & -58.8 & -47.7 & 4.9 & 4.0 &
2.6 & 2.2\\
Carina & 4  & 2.3 & 1.6 & -17.0 & -9.4 & 1.4 & 0.8 & 
0.9 & 0.5 \\
Phoenix & 33 & 3.4 & 2.7 & -33.9 & -23.2 & 2.8 & 1.9 &  1.6 & 1.2 \\
Leo I & 5  & 5.2 & 4.7 & -66.7 & -56.2 & 5.6 & 4.7 & 
2.6 & 2.3 \\
Leo II & 11 & 3.3 & 2.6 & -31.1 & -20.7 & 2.6 & 1.7 & 1.0 & 0.8\\
Aquarius & 9  & 3.5 & 2.8 & -34.5 & -23.8 & 2.9 & 2.0 & 2.5 & 1.9 \\
Sextans & 45  & 2.8 & 2.1 & -23.6 & -14.5 & 2.0 & 1.2 &  2.5 & 1.7\\
Fornax (del Pino) & 261  & 8.6 & 9.1 & -154.8 & f-168.7  & 12.9 & 14.0  &  
1.2 & 1.3\\
Fornax (de Boer) & 53  & 8.6 & 9.1 & -154.8 & -168.7 & 12.9 & 14.0  & 1.2 & 1.3  \\ \hline

\end{tabular}
\end{table*}

Can the energy generated by the explosions of SNe~II occurring at early times ($z>2$) be the main driver for removing the gas in some of the Local Group dwarf galaxies? Can it explain the lack of star formation in {\it fast} dwarfs at times more recent than 10 Gyr ago? 

To this end we consider the minimum energy required to expel the gas from the galaxy potential well as $\Delta W_{\rm gas}/2 = (W_{\rm f} - W_{\rm i})/2$, where the initial gravitational potential $W_{\rm i}$ is given by the sum of the DM halo gravitational potential and of the gas component, integrated out to the DM halo virial radius; while the final gravitational potential $W_{\rm f}$ is given only by the DM component, since the gas has been blown out. We also make the simplifying assumption that the density distribution of the gas follows a NFW profile as the DM halo and that the initial mass in gas is equal to the cosmological baryon fraction ($f_{\rm b} \simeq 1/6$) times the DM halo mass. Given the uncertainties in the initial amount of gas present in these systems, this appears as a reasonable first order approximation. 

We start by considering the limiting case that all the energy produced couples to the gas ($\epsilon_{\rm SN}=1$), which can be interpreted as a strong upper limit on the capability of internal feedback to remove gas at early times. In this case, we find that practically all the systems (except maybe And XVI) would produce feedback energy in large enough amount as to remove the gaseous component. This is not realistic since in our sample there are galaxies with extended star formation histories, indicating that the actual efficiency is $<1$. 

We can constrain $\epsilon_{\rm SN}$ using the fact that {\it slow} dwarfs must have hold on to their gaseous component more recently than $z=2$, to explain their extended SFHs. In particular, Fornax has a sizeable intermediate-age (1-8 Gyr old) component and has formed stars until very recently, $\sim$50-100 Myr ago \citep{coleman}, suggesting that it must have retained a large fraction of its gas more recently than 10 Gyr ago. For this dwarf, $\epsilon_{\rm SN}$ can not be higher than 10\% in order to retain the gas. Phoenix and LGS~3, which are both {\it fast} dwarfs but that still have some gas at present and have had some star formation after $z=2$ \citep{hi9}, would limit the efficiency to $\epsilon_{\rm SN} \lesssim 10\%$.

If we therefore assume $\epsilon_{\rm SN} \lesssim 10\%$, we obtain that the most luminous {\it fast} dwarfs (And~II, Cetus, Tucana, Sculptor and Sextans) could have been deprived of their gaseous component by stellar feedback.  Under our hypotheses, it is natural to expect the systems that have produced the largest $M_{\star, z>2}$ to have a positive balance between $E_{\rm SN, gas, z=2}$ and the gravitational potential: while the energy produced by SNe~II depends linearly on the stellar mass formed up to $z \simeq 2$,  the gravitational potential depends ultimately on the halo mass, which according to Tab.~\ref{tab:t2} is rather similar ($\sim 10^{9-10} M_{\odot}$) for the systems in our range of stellar masses. 

Draco and Ursa Minor, which are found well within the virial radius of the MW and have stopped forming stars by $z=2$, would need $\epsilon_{\rm SN} \simeq 15\%$ to have had their star formation quenched by stellar feedback. If, as discussed above, $\epsilon_{\rm SN}$ is likely to be lower, these simple calculations would support the possibility discussed in the previous section that other factors, like tidal and/or ram-pressure stripping from the MW and re-ionization, can either be mainly responsible for the quenching of these two galaxies, or couple to internal feedback to make gas removal easier (see also \citealt{kazantzidis}, \citealt{tomozeiu16a}, \citealt{tomozeiu16b}). On the other hand, should the DM halo masses of these galaxies be lower (as e.g. predicted by revised abundance matching relations that correct for reduced stellar mass due to quenching, e.g. Read et al. in prep.), the capability of internal feedback to remove the gas would be enhanced.

However, in this approximation we are considering that all the energy produced by the SN~II events occurring prior to $z=2$ is injected at once in the ISM. Simulations suggest that star formation proceeds in an oscillatory way, with 50-100 Myr long bursts, followed by similarly long quiescent periods (\citealt{marcolini}, \citealt{revaz09}). Part of the gas heated in a single short burst will cool down and fall back, to form the next generation of stars. So likely the net effect will be milder than what we are considering here. Clearly, the results concerning each specific galaxy should be taken with a grain of salt, but in general the comparison of the SNe energy budget with the gravitational potential at $z=2$ indicates that internal feedback alone might not have been sufficient to deprive {\it fast} dwarf galaxies from their gas component at early times, under resonable conditions of efficiency.

Even though we deem this hypothesis unlikely for the ``classical" dwarf galaxies, as those in our sample, in the Appendix we explore the impact of assuming that all the early star formation would occur by $z=6$, rather than $z=2$; we also make some consideration on feedback-driven gas loss in fainter systems than those considered here, such as those typically named as ultra-faint dwarf galaxies.

\subsection{Cuspy to cored profiles} \label{sec:cores}
\begin{figure*}
\includegraphics[width = 1.4\columnwidth]{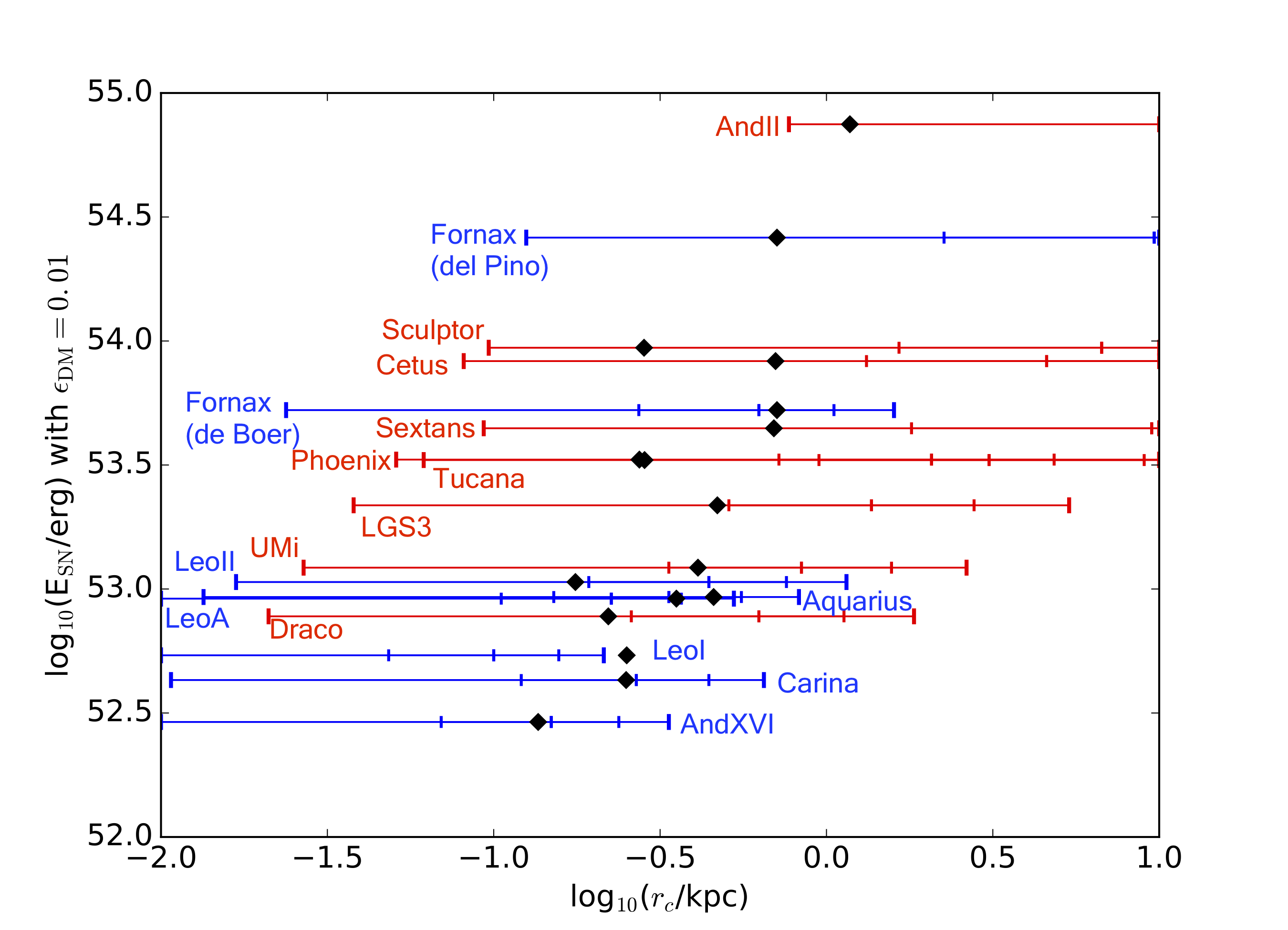}
\caption{Expected limits on the core radius at $z = 2$ for the 16 dwarfs analyzed in this work assuming the AM relation from B14. The red and blue segments represent the fast and slow dwarfs, respectively. The outer caps of the segments correspond to efficiencies of $\epsilon_{\rm DM}$ = 0.01 (left) and $\epsilon_{\rm DM}$ = 0.40 (right), and the inner caps to  $\epsilon_{\rm DM}$ = 0.1, $\epsilon_{\rm DM}$ = 0.2 and $\epsilon_{\rm DM}$ = 0.3. The black diamonds correspond to the 2-D half light radii $R_{1/2}$ of each galaxy.}
\label{fig:cores}
\end{figure*}

We explore the possibility that the early SN feedback would produce changes in the DM halo density profile of the dwarf galaxies. It has been discussed, e.g., in \citet{penarrubia}, that the transformation from cuspy, NFW profiles to cored profiles requires such an amount of energy that can only be generated by SNe~II explosions. We follow these authors' formalism and consider the 3D mass density of the DM halo to follow the profile:
\begin{equation}
\rho_c (r) = \frac{\rho_s r_s^3}{(r_c+r)(r_s+r)^2}
\label{c}
\end{equation}
where $r_c$ is the core radius, and $r_s$ and $\rho_s$ are the charasteristic inner radius and density of the NFW profile. It can be shown that Eq.~(\ref{c}) reduces to the NFW profile described by equation (\ref{nfw}) when $r_c = 0$. According to \citet{penarrubia}, and based on the virial theorem, the minimum energy required to core a profile is given by $\Delta W_{\rm core}/2 = (W_{\mathrm{core}} - W_{\mathrm{cusp}})/2$, where $W_{\rm core}$ is the gravitational potential of the cored DM halo, while $W_{\mathrm cusp}$ of the initially cusped DM halo. 

Given that the gravitational potential of the cored profile depends on the core radius, we calculate the minimum energy to core the profile, $\Delta W_{\rm core}/2$, as a function of the core radius $r_c$. In Tab. \ref{tab:t2} we list the value of $\Delta W_{\rm core}/2$ for creating a core of equal size to the 2-D half-light radius of each galaxy ($r_c = R_{1/2}$).  
Figure~\ref{fig:cores} shows the expected limits on the core radius that the  galaxies in our sample can form using the energy from the SNe feedback due to $M_{\star,z>2}$ for the efficiency limits of  $\epsilon_{\rm DM}$ = 0.01 (leftmost cap in the Figure) and $\epsilon_{\rm DM}$ = 0.40 (rightmost cap in the Figure) and assuming the B14 AM relation; the results for the M13 AM relation are extremely similar. 

For each given galaxy, the results are heavily dependent on the choice for $\epsilon_{\rm DM}$ as there can be up to two orders of magnitude difference between the values of core radius obtained with $\epsilon_{\rm DM}$ = 0.01 and 0.40 (our choice of 6.5 $M_{\odot}$ as the mass limit to form a SN II has a smaller effect on the results; increasing this to 8 $M_{\odot}$ results in cores with a 60-70\% smaller $r_c$). This clearly highlights the need for more constrained estimates of the efficiency parameter. 

If the main mechanism for core formation is the response of the DM halo to the change of gravitational potential induced by repeated short bursts of star formation, as suggested by simulations (e.g. \citealt{read05}, \citealt{pontzen}), then the amount of SN~II energy that radiates away does not contribute to this process, and therefore $\epsilon_{\rm DM}$ can not be larger than $\epsilon_{\rm SN}$. If we consider the constraints on the efficiency discussed in Sect.~\ref{sec:gas}, then we obtain $\epsilon_{\rm DM} \lesssim$ 10-15\%. 

For both the B14 and M13 AM relations, the galaxies with the largest $M_{\star,z>2}$ produce $r_c > 5$ kpc already with an efficiency of $\epsilon_{\rm DM} \sim$ 0.1-0.2.
Observationally, there is an on-going debate as to whether MW satellites of the type considered here inhabit cuspy or cored DM haloes (e.g. \citealt{goerdt06}; \citealt{bat08}; \citealt{walker11}; \citealt{agnello}; \citealt{cole12};
\citealt{richardson14}; \citealt{strigari}; \citealt{jardel}; \citealt{breddels14}). The observed kinematic (and light distribution) properties of the stellar component  --and globular cluster system, for Fornax-- are consistent with core sizes of $\sim$0.5-2 kpc. The upper limit on the core size is typically difficult to constrain (see e.g. \citealt{penarrubia}), however \citet{amorisco} exploited the presence of three stellar components among the Fornax red giant branch stars to limit the core size\footnote{We note that the authors do not statistically exclude a cuspy, NFW halo.} of Fornax DM halo to $1.0_{-0.4}^{+0.8}$ kpc. In this respect, core sizes $r_c > 5$ kpc can be considered as unlikely to be realistic, and this would be telling us that the $\epsilon_{\rm DM}$ ought to be less than 0.2. This is compatible with the constraints on the efficiency we provide in Sect~4.2.

Except for Fornax when using the SFH from \citet{pino13}, all the {\it slow} dwarfs should have difficulties to form a core of $r_c \gtrsim$ 1 kpc given the conditions at $z = 2$ and would need a larger $\epsilon_{\rm DM}$ than what we constrain here to be able to produce core radii as large as their half-light radius by $z = 2$. On the other hand, the {\it fast} dwarfs have on average larger $M_{\star, z>2}$ and corresponding injection of energy into the ISM: in the regime of $\epsilon_{\rm DM}\lesssim 0.1$, the {\it fast} dwarfs appear to be able to form a core of $r_c \sim R_{1/2}$ by $z = 2$, unless the efficiency is of only a few percent. This could suggest that cuspy profiles (or the smallest cores) should be found preferentially in {\it slow} dwarfs, and it is in agreement with previous results from the analysis of \citet{brook15} (see their Fig. 7). In their work, two additional dwarfs are expected to be cuspy, namely Draco and UMi. They are the only 2 {\it fast} dwarfs in our scheme that form less than $10^6$ M$_{\odot}$ in stars before $z=2$, making the creation of a large core more difficult.

In our calculations, however, we are focusing on the energy budget at $z \sim 2$, ignoring stellar feedback from subsequent star formation. Hydrodynamical simulations by \citet{read} and \citet{dicintio17} show that cores with sizes comparable to the 3D half-light radius\footnote{\citet{wolf} show that the 3D half-light radius is $\sim$1.3 the 2D half-light radius for a variety of commonly used surface density profiles.} of the stellar component can eventually form if star formation proceeds long enough (depending on the DM halo mass vs energy that couples to the DM). Subsequent star formation would have to compete against a growing DM halo, which might slow down the increase in core size, an effect that cannot be estimated in the simulations by R16 since the DM halo of their simulated dwarfs is not growing in time; however it is probably safe to consider the core sizes of {\it slow} dwarfs as lower limits. On the other hand, the SF of {\it fast} dwarfs is either completely halted by $z=2$ or just had some small residual activities: therefore we do not expect an additional growth of the DM halo core size due to this effect.

As discussed by \citet{maxwell15}, the \citet{penarrubia} approach has built in the assumption that the DM mass from the innermost regions can be redistributed by the feedback all to way to the DM halo virial radius. This leads to a larger amount of energy with respect to the one required to redistribute the DM halo mass within the region of the core, making core formation more difficult. The \citet{penarrubia} and \citet{maxwell15} approaches agree in the regime of 2-4 kpc core radii for DM halo masses in the range considered here. On the other hand, for smaller core radii, the estimates we are providing with the \citet{penarrubia} approach should be regarded as lower limits. Therefore the formation of cores via internal feedback appears to be energetically feasible in the range of stellar and halo masses here considered.

We note that the B14 and M13 AM relations produce among the largest values of DM halo masses associated to dwarf galaxies of a given luminosity, which for the dwarf galaxies here analyzed range between $1.5-10 \times 10^9$ M$_{\odot}$ at $z=2$. If we evolve back to $z=2$ the DM halo masses predicted for the AM relation by \citet{behroozi}, which is the most different from the previous two, the DM halo masses would be much smaller, ranging from $\sim 1.5 \times 10^8 M_{\odot}$ to $\sim 3 \times 10^9 M_{\odot}$, yielding even larger core radii (see  Appendix~\ref{sec:lower}). 

Given the exquisite spectroscopic data-sets of several hundreds, or even thousands, accurate l.o.s. velocities of individual stars existing for the bright early-type  galaxies satellites of the Milky Way (e.g. see review articles \citealt{bat13}, \citealt{walker}), the exciting recent measurement of the internal transverse motion of one of them \citep{massari}, and the progress in sophisticated dynamical modeling tools (e.g. \citealt{breddels13}, \citealt{zhu}, \citealt{read17}), MW satellite galaxies are the Local Group dwarfs for which we can aim to have the best DM halo properties determinations. Among the MW satellites in our sample, those expected to have still a cusp or the smallest cores are Draco and Ursa Minor, while the largest ones are likely to be found in Sculptor and Fornax.

Note that \citet{laporte} discuss that the accretion of dark haloes can result into a cusp re-growth, however it is unclear on which timescales and what environments this process is more likely to occurr.  

\section{Summary and conclusions}
\label{sec:conclusions}

In this paper we have performed an observationally motivated analysis of the early evolution of 16 Local Group dwarf galaxies, using accurate SFHs from the literature. We follow the classification in {\it fast} and {\it slow} dwarfs proposed by \citet{gallart} and study whether their different present-day properties and life-time evolution can be traced back to differences in the early properties of these two main galaxy types.

Since the SFHs of our sample of dwarfs are usually derived from photometric data-sets that cover only a fraction of the dwarfs' stellar component, we correct for the incomplete spatial sampling using statistical tools. To this end, we create mock galaxies following the surface density profile and structural parameters of the ancient stars ($>$10Gyr old, i.e. formed prior to $z=2$) stars. The information on the spatial properties of the ancient stars is obtained either from the literature, when available, or by our own MCMC analysis of horizontal branch stars selected from wide-area photometric catalogues. We integrate the SFHs up to $z = 2$ ($\sim$ 10 Gyr ago) and correct the resulting formed stellar mass for the missing coverage. Our correction is found to be non-negligible in the majority of the cases.

We find that {\it fast} dwarfs formed more stellar mass by $z=2$ than {\it slow} types over the 2 orders of magnitude probed by the data. This result adds information in absolute terms that was missing in the relative comparison by \citet{gallart}. The availability of more SFHs in the literature for dwarfs of larger stellar mass, like WLM, could confirm if this trend holds also at present-day stellar masses $> 10^7 M_{\odot}$. Additionally, we find hints that the DM haloes of {\it fast} dwarfs have on average a larger dynamical mass than those of {\it slow types} within the half-light radius. We also find a correlation between the dwarfs' dynamical mass within the half-light radius and the amount of stars formed by $z=2$, which is clearer than when considering instead their present-day luminosity; we interpret this as $M_{\star,z>2}$ being a better indicator of the initial relative baryonic content of the galaxies in the sample, before environmental effects might have deprived some of them of their gaseous component (and prevented subsequent growth in stellar mass).  

Our estimation of the stellar mass formed up to $z=2$ is also useful to explore if stellar feedback could have removed the gas component of the {\it fast} dwarfs and to what extent it might have caused a transformation from a cuspy to cored DM halo. As expected, a key, but unknown, parameter in this kind of estimates is the efficiency with which the SN energy couples to the gas and DM. By requiring that dwarfs that have experienced significant star formation more recently than $z=2$ cannot have been deprived of their gaseous component at ancient times, we are able to put limits on the possible amount of gas coupling efficiency, and consequently on the capability of feedback to halt star formation in {\it fast} dwarfs and to core DM profiles. Our limits are compatible with the observational constraints on the core radius estimates for the Fornax dwarf galaxy.

We find that the gas removal by $z=2$ driven only by internal feedback would be possible in the {\it fast} dwarf galaxies with a massive stellar component, under reasonable conditions of efficiency according to our limits ($\epsilon_{\rm SN} \lesssim 10-15\%$). Our analysis however assumes that all the SN~II energy is injected at once, rather than in short bursts followed by quiescient periods, which might overestimate the capability of the feedback to expel the gas. Therefore it is more likely that internal feedback alone cannot explain the quenching of star formation in {\it fast dwarfs}, in particular at the fainter end, and that other factors, such as ram/tidal-stripping and/or re-ionization might play a role too. 

Regarding the `cusp-core' problem, we find that the feedback energy would have been enough to produce a transformation from cuspy to cored profiles in most of the dwarf galaxies in the sample by $z=2$. Our result is quite degenerated depending on the assumed feedback efficiency. This parameter is one of the most important quantities to be constrained in order to break the degeneracy with the core radius size. For the range of efficiencies that we can constrain using the fact that {\it slow } dwarfs cannot have removed their gaseous component, we find that {\it fast} dwarfs could have formed a core of size of the order of their 2-D half-light radius ($r_c \sim R_{1/2}$) by $z=2$.  

The dark matter core sizes we derive here should be considered as lower limits: we neglect the feedback after $z=2$, which, if considered, would yield larger cores for the {\it slow} dwarfs; and as discussed in Sect.~\ref{sec:cores} the \citet{penarrubia} formalism has higher energy requirements for core formation with respect to the \citet{maxwell15} formalism. However, unless the SN~II energy coupling to the DM is of only a few percent, cores of at least 0.1-0.2 kpc appear to be energetically feasible to produce even within our conservative approach.

Among the MW satellites considered here, the systems which offer the best prospects for detecting a cusp (or where we expect the smaller cores to be) appear to be systems such as Draco and Ursa Minor, which also have {\it fast} SFHs and therefore do not suffer from the possibility of core size increase due to neglected SF $<$10 Gyr ago. At the opposite end sit Sculptor and Fornax (with the \citealt{pino13} SFH); in particular the latter, given its significant SF at intermediate ages, should be the system most likely to host a comparatively large core. We emphasize that the ``core-most" and ``cuspy-most" MW satellites  are the same ones that were  obtained with a different method in \citet{brook15}.

\section*{Acknowledgements}
The authors are grateful to S. Hidalgo for providing the estimate of the half-light radius of the ancient component of Leo~A, T. J. L. de Boer for the determination of the SFH of Fornax, P. Stetson and M. Irwin for kindly sharing reduced photometric catalogues and the anonymous referee for useful suggestions.
JRBC gratefully acknowledges Instituto de Astrof\'isica de Canarias (IAC) for its hospitality and travel support during visits. GB  gratefully acknowledges financial support by the Spanish Ministry of Economy and Competitiveness (MINECO) under the Ramon y Cajal Programme (RYC-2012-11537) and the grant AYA2014-56795-P; the latter grant supports also CG, MM, LC. ADC acknowledges financial support from a Marie-Sklodowska-Curie Individual Fellowship grant, H2020-MSCA-IF-2016 
Grant agreement 748213, DIGESTIVO. JIR would like to acknowledge support from the STFC consolidated grant ST/M000990/1 and the MERAC foundation.




\bibliographystyle{mnras}
\bibliography{bib.bib} 





\appendix

\section{Feedback at lower DM halo masses} \label{sec:lower}

Uncertainties in the SFH determination may shift the peak of the SF to $\sim$1.2 Gyr younger ages than the true one and to artificially widen the age distribution (e.g. \citealt{aparicio16}). One might then wonder if it is possible that all the star formation activity that we have so far considered as taking place out to $z=2$ was in reality confined to higher redshift, e.g. $z=6$ (1 Gyr from the start of SF), i.e. to the pre-reionization era. At that time the DM halo would have been smaller and stellar feedback (coupled with the re-ionization UV background) could have been more efficient both in removing the gaseous component and in transforming a DM cusp into a core. 

We deem this hypothesis unlikely for the kind of dwarf galaxies we are considering in this work: (1) the presence of a "knee" in the [$\alpha$/Fe] versus [Fe/H] trend for the almost purely old {\it fast} dwarf galaxies in the sample would suggest that chemical enrichment (and star formation) have been on-going for more than 1-2 Gyrs in these galaxies (see \citealt{boer12} for an age dating of the "knee" in Sculptor). (2) \citet{bovill} show that it is unlikely that dwarf galaxies brighter than 1 million $L_{V, \odot}$ have formed more than 70\% of their stars by $z=6$. By analyzing the SFH of Cetus, Tucana, LGS~3 and Phoenix (3 of which are fainter than $L_{V, \odot} = 1 \times 10^6 $) taking into account uncertainties in SFH determination, \citet{aparicio16} conclude that also these galaxies are unlikely to be reionization fossils. (3) From the work of \citet{sawala16}, it appears likely that dwarf galaxies as luminous as those considered here started and/or continued forming stars after re-ionization was completed. 

Nonetheless, we will still revisit the calculations performed in the previous sections to understand the impact on our conclusion that stellar feedback cannot be the main reason for halting SF in most {\it fast} dwarfs; this also give us the opportunity to compare our predicted core sizes to those of R16, because the smaller halo masses predicted at $z=6$ by the B14 and M13 relation are more similar to those considered in the hydrodynamical simulations by \citet{read}.  

In order to determine the gravitational potential at $z=6$, we follow the same procedure as in Sect.~4, but considering that at $z=6$ the ratio between $M_{\mathrm{vir}} (z=6)$ and $M_{\mathrm{vir}} (z=0)$ according to \citet{fakhouri} is about a $\sim$10\% for haloes in our mass regime ($\sim 10^{10} M_{\odot}$), and using the parameters of the equation (\ref{cvir}) for $z = 6$ from \citet{dutton}. This results in DM halo masses ranging between $5 \times 10^8 M_{\odot}$ and $1.3 \times 10^9 M_{\odot}$. In this sense, the effect of making the calculations with the DM halo mass at $z=6$ for the B14 and M13 AM relations is similar to choosing at $z=2$ the AM relation by \citet{behroozi} (which yields DM halo masses that range from $\sim 1.5 \times 10^8 M_{\odot}$ to $\sim 3 \times 10^9 M_{\odot}$ for our sample).

$\bullet$ Using the DM halo masses predicted at $z=6$ with the B14 and M13 AM relations, we find that for  $\epsilon_{\rm SN}$ = 0.1 all the galaxies could remove their gas, including the {\it slow} types, which appears unrealistic since {\it slow} types have formed stars for practically a Hubble time and therefore they must have been able to hold on to a gas reservoir. We find that the maximum efficiency compatible with the slow dwarfs not losing their gas by $z=6$ would be $\epsilon_{\rm SN} \lesssim 0.5\%$. Considering that $\epsilon_{\rm DM} \leq \epsilon_{\rm SN}$, this is compatible with the expected limits on the core radii of $\sim$ 2-5 kpc.

As mentioned above, the \citet{behroozi} AM relation at $z=2$ yields comparable DM halo masses to those from the B14 and M13 relation at $z=6$. If we evolve back the DM halo masses with the \citet{behroozi} AM relation at $z=6$, the behaviour in terms of expected core radii would be even more catastrophic, which would set the efficiency then to even lower values. 

$\bullet$ The results from our simple calculations are in good agreement with the outcome of hydrodynamical simulations by \citet{read}, where the spatial scales relevant to follow the impact of individual SNe events are resolved. Let us for example focus on the case of And~XVI and Leo~A, which according to our calculations have a DM halo mass of $5\times10^8 M_{\odot}$ and $\sim 1 \times 10^9 M_{\odot}$ at $z=6$. These numbers are directly comparable to the $5\times10^8 M_{\odot}$ and $\sim 1 \times 10^9 M_{\odot}$ DM haloes in \citet{read} (hereafter, medium and large R16 DM halo). Those simulated haloes formed a stellar component of $M_{\rm \star,birth} = 12.6 \times 10^5 M_{\odot}$ and $7.1 \times10^6 M_{\odot}$, respectively, whose stellar feedback was found to have a coupling efficiency of 2\%, and created a core radius = 0.3 kpc in the former and 0.6 kpc in the latter, albeit after a long time (after 8 and 12 Gyr, respectively). For forming a dark matter core, what matters is the energy cumulatively injected into a DM halo of a given mass. The \citet{read} conditions in terms of injected energy in their medium and large DM halo are comparable to our $M_{\star,z>2}$ of And~XVI and Leo~A if we assume $\epsilon_{\rm DM}= 0.01$ for the former and $\epsilon_{\rm DM}= 0.02$ for the latter;  the resulting core radii are $\sim 0.1$ kpc and $\sim 0.3$ kpc, respectively. Considering the different IMF assumed and concentration parameter, the core sizes found in R16 hydrodynamical simulations and those predicted by our calculations are in good agreement. This lends support to the validity of our simplified approach.

$\bullet$ The surroundings of the large LG spirals host a wealth of much fainter systems than ``classical" dwarf galaxies, commonly called ultra faint dwarfs (UFDs). Some of these have SFHs consistent with having formed the great majority of their stars by $z=6$ (e.g. \citealt{brown}) or even being ``fossil" galaxies (e.g. \citealt{frebel}). For these systems, e.g. of ancient stellar masses $\sim 10^4$-$10^5 M_{\odot}$, the energy balance calculated considering DM halo masses at $z=6$, and requiring that the efficiency be $<10-15$\%, would result in the $10^5 M_{\odot}$ systems being able to expel gas by internal feedback alone, while external effects would need to be invoked for the even fainter ones. In terms of core formation, it appears that the amount of available SN~II feedback would be capable of forming a DM core larger than 0.1 - 0.2 kpc, i.e. of the order of the half-light radii observed for galactic systems of these stellar masses, already at extremely low efficiencies, between 1 and 10\%. The number of SN~II produced is $\sim 280$ and $\sim 2800$ in the $10^4$ and $10^5$ cases, therefore stochastic sampling of the IMF in evaluating the produced internal feedback is probably not an issue at these stellar masses \citep[see also e.g.][]{Revaz2016}. It remains to be assessed whether e.g. the assumptions of DM halo growth history and expected concentration at a given DM halo mass are appropriate also for the class of UFD systems.


\bsp	
\label{lastpage}
\end{document}